\documentclass[12pt]{elsarticle}
\def\la{\langle}
\def\ra{\rangle}

\def\be{\begin{equation}}
\def\ee{\end{equation}}

\usepackage{graphics,graphicx}
\journal{Physica A}

\begin{document}

\title
{Scaling and crossover behaviour in a truncated  long range quantum  walk}

\author{Parongama Sen}%
\address{Department of Physics, University of Calcutta, 92 Acharya Prafulla Chandra Road, Kolkata 700009, India.}

%\cu

%\documentclass[12pt]{elsarticle}
%\documentclass[aps,twocolumn,superscriptaddress]{revtex4}
%\def\la{\langle}
%\def\ra{\rangle}
%\def\ua{\uparrow}
%\def\da{\downarrow}
%\def\be{\begin{equation}}
%\def\ee{\end{equation}}
%
%
%%\usepackage{amsmath}
%%\usepackage{amsfonts}
%\usepackage{amssymb}
%
%\usepackage{graphics,graphicx}
%\journal{Physica A}
%%\usepackage{dcolumn,bm}
%
%%\newcommand{\cu}
%%{\affiliation{Department of Physics, University of Calcutta, 
%%92 Acharya Prafulla Chandra Road, Kolkata 700009, India.}}
%
%
%%\topmargin 0.01cm
%
%
%\begin{document}
%

%\author{Parongama Sen}
%%\email[Email: ]{psphy@caluniv.ac.in}
%\address{Department of Physics, University of Calcutta, 92 Acharya Prafulla Chandra Road, Kolkata 700009, India.}

\begin{abstract}
We consider a 
 discrete time quantum walker in one dimension, where at each step, the 
step length  $\ell$ is chosen from a distribution 
$P(\ell) \propto \ell^{-\delta -1}$ with $\ell \leq \ell_{max}$.
We evaluate the probability $f(x,t)$ that the walker is at position $x$ at time $t$ and its first two moments.
As expected,  the disorder   effectively localizes the walk  even for large values of $\delta$. 
Asymptotically,  $\langle x^2 \rangle \propto t^{3/2}$ and $\langle x \rangle \propto t^{1/2}$ independent of $\delta$ and $\ell$, both finite.
The scaled distribution   $f(x,t)t^{1/2}$ plotted   versus $x/t^{1/2}$ 
shows a data collapse for $x/t <    \alpha(\delta,\ell_{max}) \sim \mathcal O(1)   $ indicating the existence of a  universal  scaling function. The scaling function is shown to have a crossover behaviour 
 at $\delta  = \delta^* \approx 4.0$ beyond which the results are independent of $\ell_{max}$. 
We also calculate  the von Neumann entropy 
of entanglement which gives a larger asymptotic value 
compared to the quantum walk with unique step length even for large $\delta$, with negligible
dependence on the initial condition.

\end{abstract}

\maketitle

\section{Introduction}

Discrete time quantum walks on discrete space have been studied extensively over the last couple of decades \cite{Aharonov,kempe,nayak,amban}. 
In the quantum walk in one dimension, the state
of the walker is expressed in the $|x\rangle \otimes |d\rangle$ basis, where $|x\rangle$
is the position (in real space) eigenstate and $|d\rangle$ is the
chirality  or coin eigenstate (either left ($|L\ra$)  or right ($|R\ra$)).
At each time step, there is a rotation and a translation operation on the walk
which spreads according to the initial condition.
 The time dependence of  the square of the displacement for a quantum walk is $\la x^2 \ra \propto t^2$  showing it is  much faster than the classical walk (where $\la x^2 \ra \propto t$), and hence can play a key role in  many
dynamical  processes.

Various studies have also been made by  modifying  the walk in suitable ways.
In a  large proportion of these works, the effect of disorder has been studied \cite{kendon}. 
In some recent works,  the disorder has been incorporated in the   quantum walk in one dimension
by considering  long range step lengths chosen randomly at each discrete  time step \cite{ps2018,sreetama}. Variable range step length in a discrete quantum walk along with memory 
effects was also studied in \cite{elephant,elephant2}.

With a binary choice of step lengths, 
the main result  was the observation that there is a  
  sub-ballistic but super-diffusive scaling for the second moment;     $\la x^2 \ra  \propto t^{1.5}$ 
asymptotically \cite{ps2018}. 
%Interestingly, the same scaling behaviour 
%was also obtained for the quantum walk with memory. 
%In this paper, we consider a long range  walk where the step lengths are 
%chosen from a power law distribution as in a Levy walk. 
%There is a truncation in the distribution as the step lengths cannot exceed a preassigned maximum value. 
%%We are interested in further checking the universality of the scaling exponent
%of $\langle x\rangle $ and $\langle x^2\rangle $. 
%In addition, the effect of the two parameters $\delta$ and $\l_{max}$  is investigated. 
%In \cite{pototcek,ps2018}, even the choice of just two step lengths chosen randomly  
%was seen to 
%be sufficient 
%to  effectively modify  the scaling behaviour of the moments $\la x \ra$ and $\la x^2 \ra$.   
%The scaling behaviour was found to be  independent of the choice of the parameters controlling the 
%probability distribution of the step lengths in \cite{ps2018} where  binary values of step lengths was used.
%. 
In \cite{sreetama},  Poissonian and other exponentially decaying 
distributions for the step lengths were used and a similar scaling  was found (the exponent 
was reportedly $\sim 1.4$ obtained from a short time range). 
The scaling of the variance of $x$   is found to be parameter dependent for the non-Markovian walks considered in 
\cite{elephant,elephant2}, with a maximum value of the exponent equal to 3. 
%In fact, in a recent    study of a   non Markovian quantum walk 
%\cite{elephant},  which essentially used long ranged 
%steps, the scaling $\la x^2 \ra \propto t^{3/2}$ was shown analytically. 
%However, in this work,  a long term memory effect was also included.   
In this context it may be mentioned that long ranged jumps were also 
considered albeit in a different manner, arising due to reasons like  inhomogeneity of the material 
or geometry of the interferometer
\cite{potocek}.
The scaling behaviour in this case was found to be  completely different.
However, as  this long ranged walk does not belong to the class considered in  \cite{ps2018,sreetama,elephant,elephant2}, it is not relevant for the present  work
and will not be discussed further.
It may also be mentioned  that when decoherence is induced by 
periodic measurements, the walk may be regarded as one with  
random  step lengths \cite{yutaka}. However, this   walk, 
characteristed by a time dependent periodicity,
 also belongs to a different class of problems and   
  shows a ballistic to diffusive 
crossover.

%In \cite{ps2018},  based on the scaling behaviour of the probability 
%distribution $f(x,t)$ as a function of $x$ and $t$,  it was possible to define two so called decoherence parameters which showed a dependence  on the system parameters. As expected, 
%these parameters  vanish in the limit of the case without disorder. 
%
In this paper, we choose the   step lengths $\ell$ 
 from a distribution 
\begin{equation}
P(\ell) = A \ell^{-1-\delta},
\label{eq:prob}
\end{equation}
%$P(\ell) \propto \ell^{-1-\delta}$ 
which  is fat tailed. 
Moreover, we impose a cut-off in $\ell$ such that it is like a truncated Levy walk. 
Hence  $A$,  the normalization constant, is given by 
\[
A \sum_{1}^{\ell_{max}}   \ell^{-1-\delta} = 1.
\]
Note that here  the distribution is not truncated at a  particular value as in \cite{sreetama}, but rather,  the maximum step length $\ell_{max}$ is kept fixed. 
Thus  there are  two independent parameters of this distribution,  namely,  $\delta$ and $\ell_{max}$. 
 The case considered in \cite{ps2018} thus becomes a special case 
of the distribution (\ref{eq:prob}) by putting $\ell_{max} =2$  and the 
probability   $P(\ell =1)$ or $P(\ell = 2)$ can then be  expressed in terms of $\delta$.  Here, we have used $\delta \geq 0$ in general which implies larger step lengths are less probable.  

In quantum transport phenomena such  
 Levy distributions have been considered in several earlier works.  For  
the  tight binding model which includes 
interaction with a thermal bath of oscillators 
 \cite{caceres}, the effect of  long range hopping has been studied. Also, continuous time quantum walks have been considered
on discrete one dimensional lattices
with long range steps (with a cut-off),  where the 
interaction strength in the Hamiltonian decays  as a power law with the step length \cite{blumen}. 
In \cite{Chotto}, both long range hoppings and long range interactions were 
incorporated for hard core bosons.

The quantum walker with disorder (QWWD henceforth) may be slower than the quantum walker  without disorder (QWWOD henceforth) as already revealed in the earlier works, 
%for the discrete time quantum walker on one dimensional lattices, 
but because of the longer step lengths it can reach  larger 
distances.  Therefore 
 one of the  aims in studying the long ranged  quantum  walk is  to find out 
its ability to search for remote targets. 
Another is to check its dependence on the  nature of the distribution of the step lengths. 

In this paper,  we study  the scaling behaviour of $\la x \ra$ 
and $\la x^2 \ra$ 
for the fat tailed distribution given by Eq. (\ref{eq:prob}). 
Next, we focus on the form of $f(x,t)$, the probability distribution that the walker is at site $x$ at time $t$. 
Lastly, we calculate the von Neumann entropy of entanglement which develops between the position and the coin states for the long ranged walks.
 Note that  no measurement is being made  at any  time for this walk. 

\section{Dynamical scheme and results}

We generate the QWWD  in the usual manner using a Hadamard coin. 
The process is exactly the same as  described  in \cite{ps2018}
with $\ell$ chosen randomly at every step following Eq. (\ref{eq:prob}).
The state   of the particle, $\psi({x},t)$, can be written  as
\begin{equation}
\psi({x},t)=
 % \begin{bmatrix}
\left[ {\begin{array}{cc}
    \psi_{L}({ x},t)\\
    \psi_{R}({ x},t) 
%  \end{bmatrix}.
\end{array}} \right]
\end{equation}

 The walk is initialized at the origin with
$\psi_{R}(0,0) = a_0,  \psi_{L}(0,0)= b_0; ~~a_0^2 + b_0^2 =1$
and
$\psi_{L}(x\neq 0,t=0) =   \psi_{R}(x\neq 0,t=0)= 0$.
We choose  $a_0 = \sqrt{\frac{1}{3}}$ and $b_0 = \sqrt{\frac{2}{3}}$ 
\cite{corr}
such that in absence of disorder, an asymmetric probability density profile is obtained  and one can study   the  scaling of both the first and second moments.

%In this case, the system without disorder can only be achieved with $\ell_{max} =1$ and obviously 
%We have evaluated $f(x,t)$  for the long ranged walk and studied the scaling behaviour of the first two moments. The effect of the two parameters $\delta$ and $l_{max}$ is studied in detail. 
%A natural question arises about the limiting values. 
%For $\delta \to \infty$ one should be able to recover the results for
%the pure case. 
%It is difficult to study the system for a very large value of $\delta$
%as the probabilities of large $\ell$ values become extremely small. 

For any value of $\delta$, for a faithful representation of $P(\ell) $,  
 one needs to generate a  sufficient number of random 
step lengths. 
This  poses a computational difficulty which increases with  larger values of  both $\delta$ and $\ell_{max}$. 
%Actually, the larger is $\ell_{max}$, the space spanned by the walker becomes  
%larger making it difficult  to handle computationally.
The number of step lengths generated is the maximum  iteration time (i.e. the time up to which the walk is generated) multiplied by the 
number of configurations.
One has to   keep fixed the maximum time of iteration (to restrict the 
space spanned by the walker)  
and then  for each value of  $\ell_{max}$,  the minimum 
number of configurations required for a faithful representation of the 
distribution is found out.
  This naturally  determines also the maximum value of 
$\delta$ which can be used for a given value of $\ell_{max}$. 
 Obviously, for larger $\ell_{max}$, we have to
restrict to smaller values of $\delta$. 
This paper reports the cases for  $ 4 \leq \ell_{max} \leq  12$ and 
$0 \leq \delta \leq  9.5$ (the maximum value of $\delta =9.5$ corresponding to $\ell_{max} = 4$).  The number of   iterations is $t=10000$  used for $\ell_{max} =4$  and  
 $t=3000$ for $\ell_{max} = 12$.   
The number of configurations used is   
larger for larger values of $\ell_{max}$ for the reasons stated above;  
for example, it is 
50000 for $\ell_{max} =12$    
and 10000 for $\ell_{max} = 4$.

\begin{figure}
\includegraphics[width=5cm,angle=270]{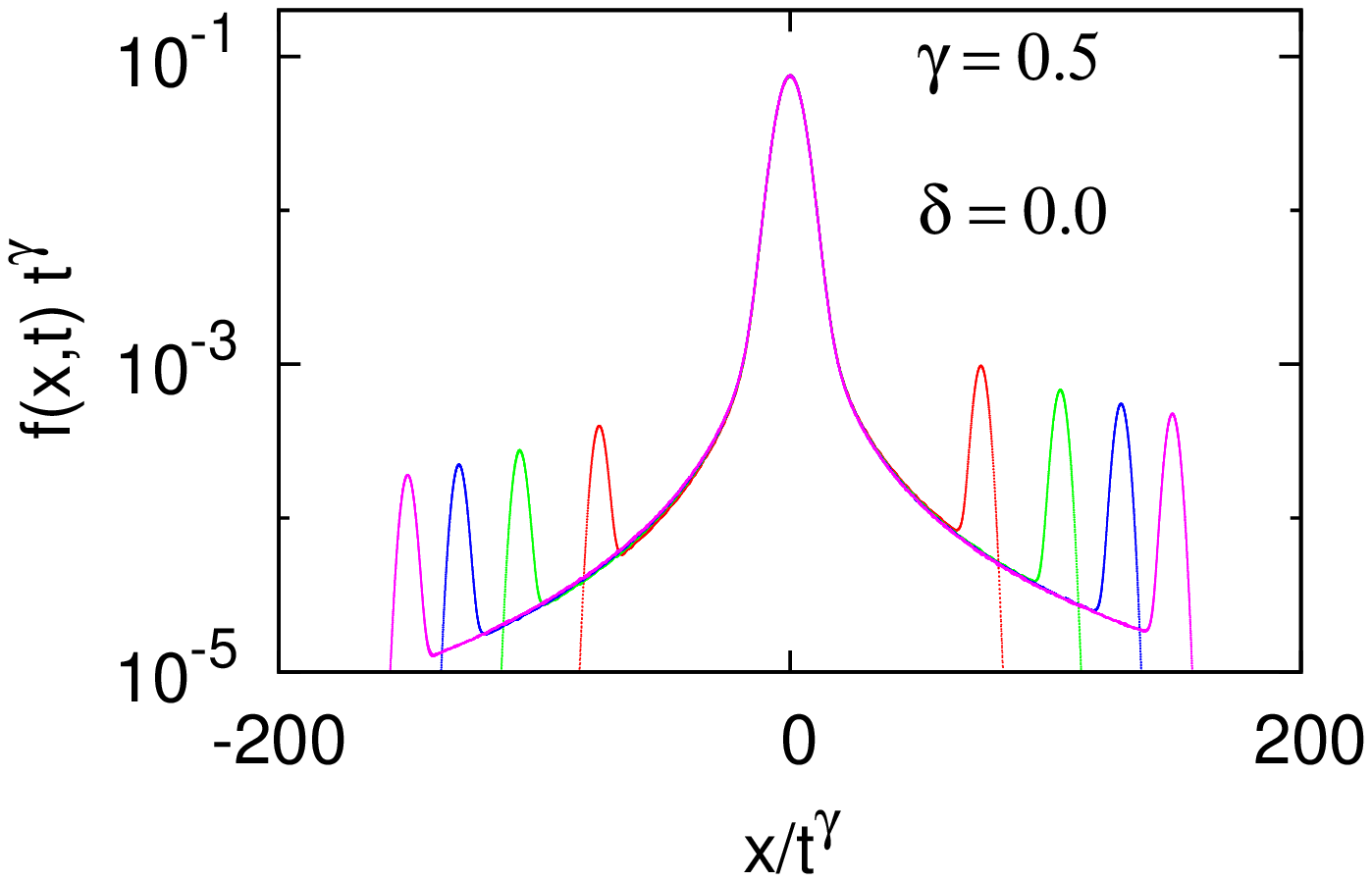}
\includegraphics[width=5cm,angle=270]{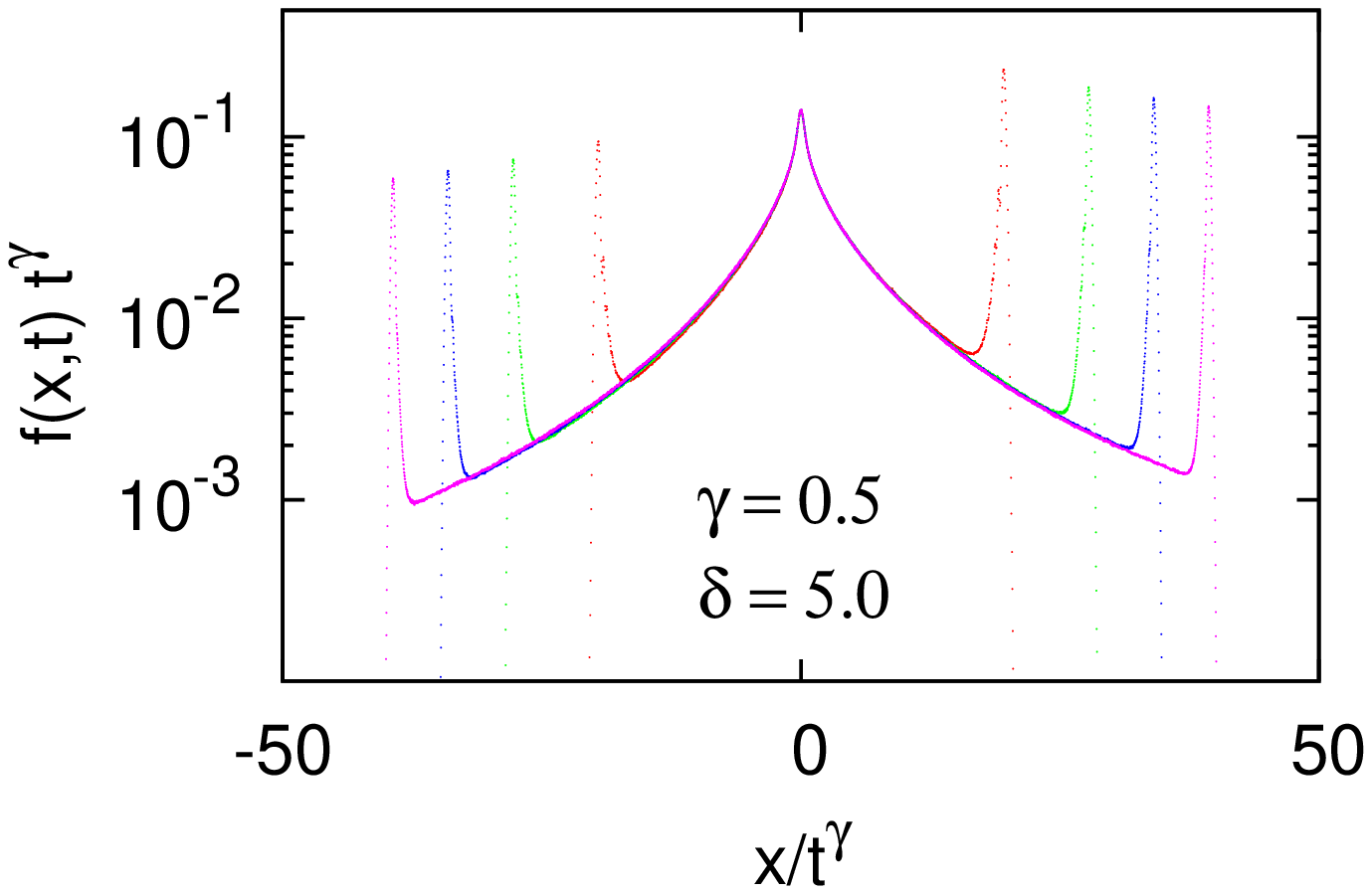}
\caption{Data collapse of $f(x,t)t^{1/2}$ against $x/t^{1/2}$  
for four different values of time $t$ shown for 
$\delta = 0$ (left) and $\delta = 5$ (right) with   $\ell_{max} = 12$.}
\label{fig1:scaling_cent}
\end{figure}

The probability $f(x,t)$ for the particle
to be at $x$ is  given by $|\psi_{L}({ x},t)|^2 + |\psi_{R}({ x},t)|^2$ and adds up to  unity for all $t$. It has typically a central peak and two ballistic peaks occurring at nearly extreme values of $x \sim t$.
%We first discuss the nature of  the distributions for $\ell_{max} = 12$  
%for two extreme values of $\delta$. 
We find that except for the region very close to the ballistic peaks, a data collapse can be obtained by plotting $f(x,t)t^{\gamma}$ against $x/t^{\gamma}$ with $\gamma = 0.5$.
%On the other hand, unlike the case for $\ell_{max} =2$, already discussed in \cite{ps2018}, the data collapse for the ballistic peaks are not as 
%impressive when the data is scaled by $t$ (especially at smaller values of $\delta$); only the ballistic peak values collapse.    
Fig. \ref{fig1:scaling_cent} 
%and
%% \ref{fig2:scaling_cent} 
shows the results for the collapse of the centrally peaked region
for $\ell_{max}=12$ and two extreme values of $\delta$.

\begin{figure}
\includegraphics[angle=0,width=5cm, angle = 270]{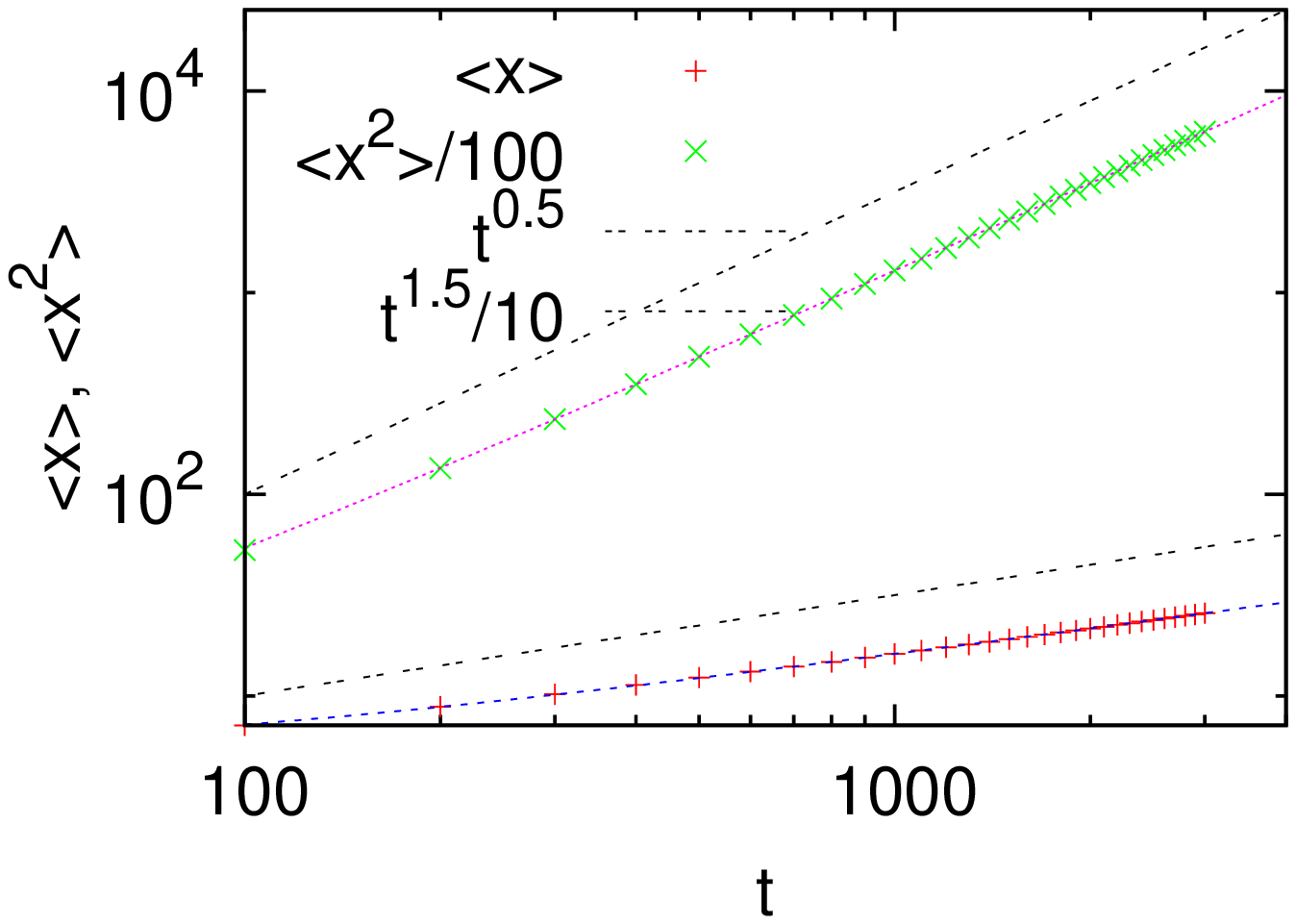}
\includegraphics[angle=0,width=5cm, angle = 270]{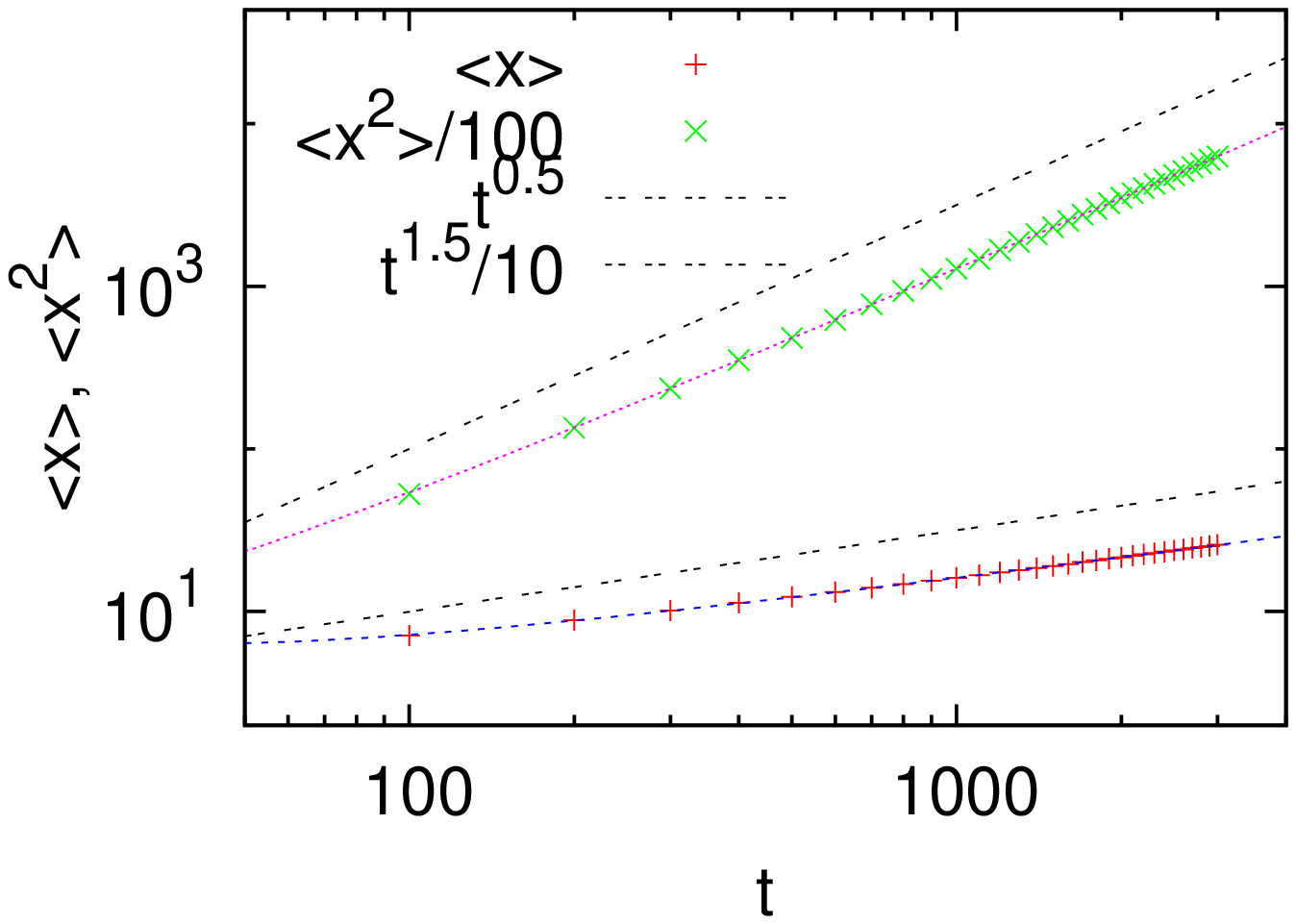}
\caption{The first two moments of $f(x,t)$ for  $\delta = 0$ (left) and $\delta = 5$ (right)  are 
shown for $\ell_{max}=12 $. The continuous lines are best fit curves obtained using the   
form  given in  Equations \ref{eq:mom1} and \ref{eq:mom2}. The dashed lines have simple power law variation 
indicated in the key.
}
\label{fig2:fits}
\end{figure}

The first two moments are evaluated  as a function of time; 
the plots are shown  in Fig. \ref{fig2:fits}  for
$\ell_{max} = 12$ and   $\delta = 0$ and 5. 
Clearly,
the asymptotic variation is $\la x\ra \propto t^{1/2}$ and 
$\la x^2\ra \propto t^{3/2}$. Following \cite{ps2018}, we attempt to fit the 
moments using the equations
\begin{equation}
\langle x\rangle = t/(b_1+{b_2}\sqrt{t})
\label{eq:mom1}
\end{equation}
and
\begin{equation}
\langle x^2\rangle = t^2/(b_3+b_4\sqrt{t}),
\label{eq:mom2}
\end{equation}
and find that a good consistency can be achieved. These equations had been arrived at using an approximate form of $f(x,t)$ assuming delta function like behaviour at $x=0$ and at the two extreme values of $x$ corresponding to the ballistic 
peaks.  This approximation may not work well here  
as the collapse for the ballistic peaks is not as accurate as in \cite{ps2018} 
when one plots  $f(x,t)t^{\gamma}$ against $x/t^{\gamma}$ with $\gamma=1$,  
especially for small values of $\delta$
(only the tips of the ballistic peaks, which have  finite widths for small 
$\delta$,  merge). 
This  is borne by the fact that one obtains negative values of $b_3$ for small values of $\delta$ which is unphysical. Hence we do not proceed with further analysis of the parameters although the 
fitting is found to be good. 
It may also be added that scaling behaviour of the moments discussed above is independent of the value of $\ell_{max}$ and $\delta$.

%The scaling behaviour is however, definitely in consistency with the earlier 
%result further confirming that it is independent of the form of the 
%distribution $P{\ell}$. 

%\begin{figure}
%\includegraphics[width=5cm,angle=270]{long_fig1_lmax12_alpha_0.eps}
%
%\includegraphics[width=5cm,angle=270]{long_fig1_lmax12_alpha_5.eps}
%\caption{Plot of  $f(x,t)t$ against $x/t$  
%for four different values of $t$ 
%$\delta = 0$  (top) and $\delta = 5$ (bottom) with  $\ell_{max} =12$ does not show 
%a good collapse although the tips of the ballistic peaks coincide.} 
%\label{fig2:scaling_cent}
%\end{figure}

\begin{figure}
\includegraphics[angle=0,width=5cm, angle = 270]{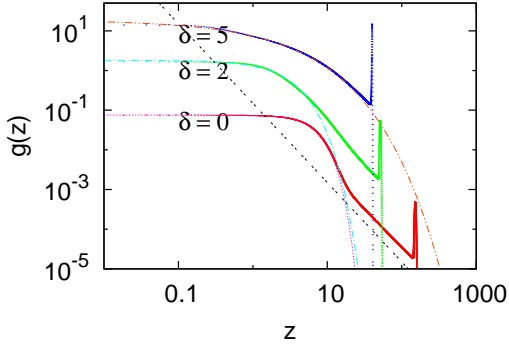}
\caption{The scaled distribution $g(z)$ fitted with stretched exponential forms for different values of $\delta$    for $\ell_{max} =12$. The data has been shifted vertically   
 for better visual effect for $\delta = 2$ and $5$. The dashed line has the form $z^{-2}$.}
\label{fig3:scaling}
\end{figure}

Next we analyze in detail  the characteristic features  and form of the distribution
$f(x,t)$. 
Note that in most of the earlier works, the disorder, usually incorporated through the coin operator,  resulted in the  transition to a  Gaussian distribution. 
In contrast, in the present  case, where the step lengths are chosen from a Levy distribution,  the form of $f(x,t)$ as a function of the space variable $x$ is non-Gaussian in general.
It is also obvious from Fig. \ref{fig1:scaling_cent} that the nature of the distribution changes appreciably as $\delta$ is changed.
In the following, we investigate the form of the centrally peaked region that is obtained
after suitable rescaling of the data.
The  scaled distribution is written as $ g(z) = f(x,t)t^{1/2}$ and is studied as a 
function of $z = x/t^{1/2}$ for the largest value of $t$ ($t_{max}$) used in the generation of the walk. 
We have observed that up to values of $x = \alpha t^{1/2}_{max}$ the data
can be fitted to a unique scaled function, where  
 $\alpha$ is dependent on $\delta$ and $\ell_{max}$.
We thus discuss the behaviour of  $g(z)$ for 
 $z < z_{max} = \alpha$. For small values of $z$,  
$g(z)$  shows a stretched exponential behaviour while we note that there is a distinct power law decay for larger values of $z$.
Thus we conclude 
the following behaviour for the scaling function  $g(z)$:

\begin{eqnarray}
g(z) & = & a \exp(-b z^c), ~~~~~    z < z^*
\label{eq:scaling}\\
& 	= & {\rm const} ~~~z^{-2}, ~~~~~   z^* < z <  z_{max}
\nonumber
\end{eqnarray}
For  values of $\delta$ greater than  $\simeq 4$, $z^*$ coincides with $z_{max}$ such that the power
law region is absent.  In Fig. \ref{fig3:scaling}, the data for three different $\delta$ values are plotted along with the best fit stretched exponential curves 
for $\ell_{max} = 12$. The function  $z^{-2}$ is shown separately. 

\begin{figure}
\includegraphics[angle=0,width=5cm, angle = 270]{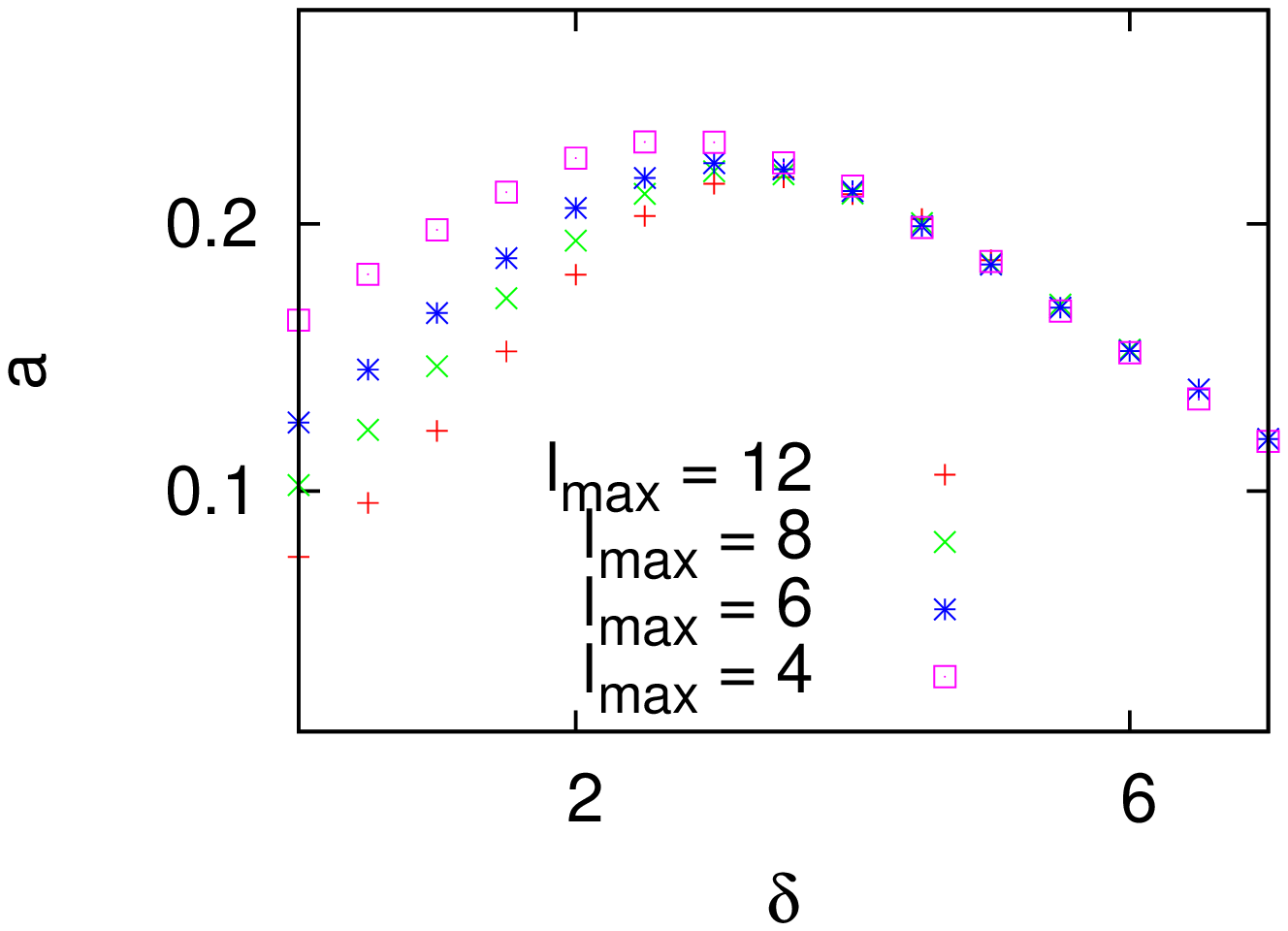}
\includegraphics[angle=0,width=5cm, angle = 270]{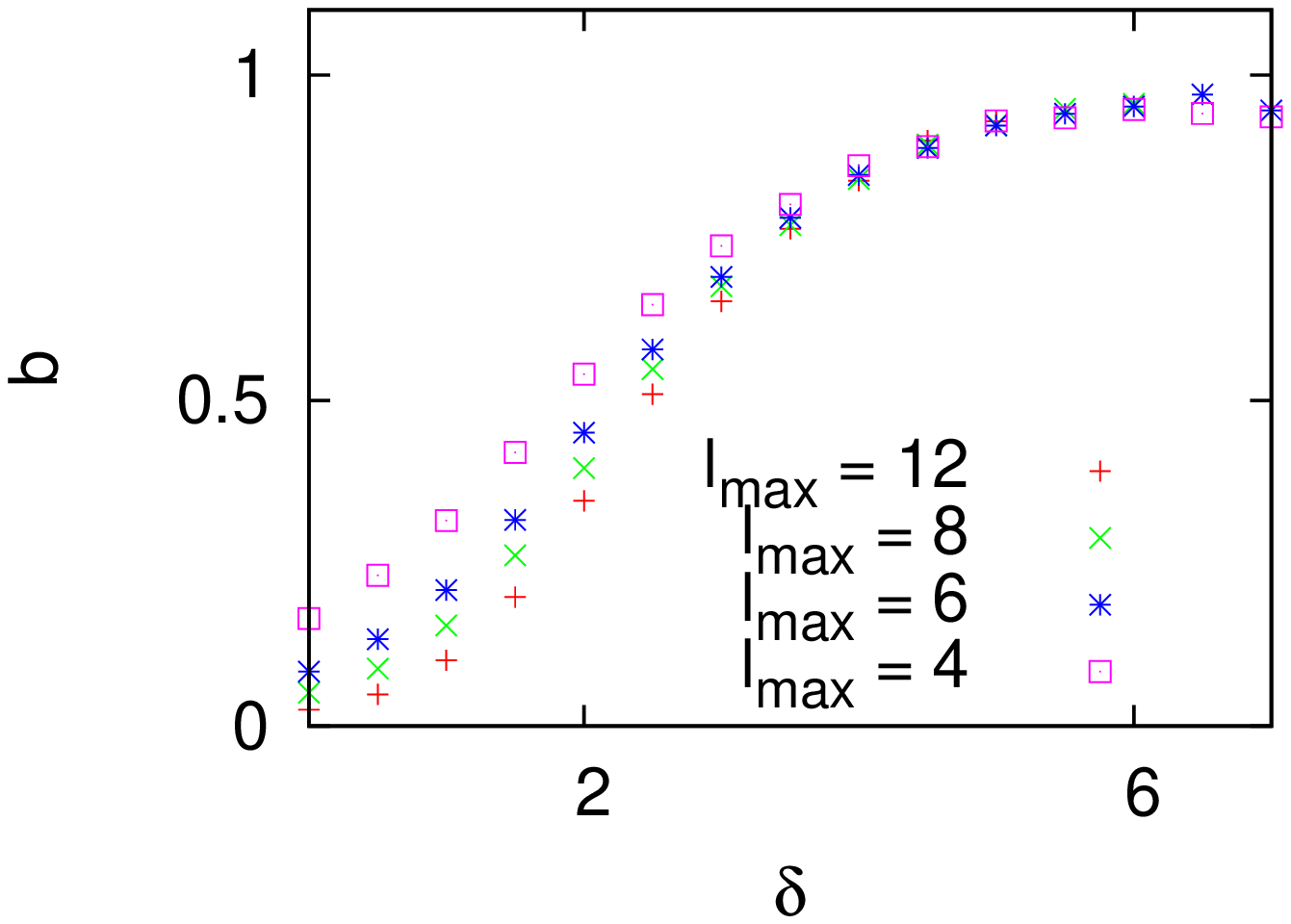}

\includegraphics[angle=0,width=5cm, angle = 270]{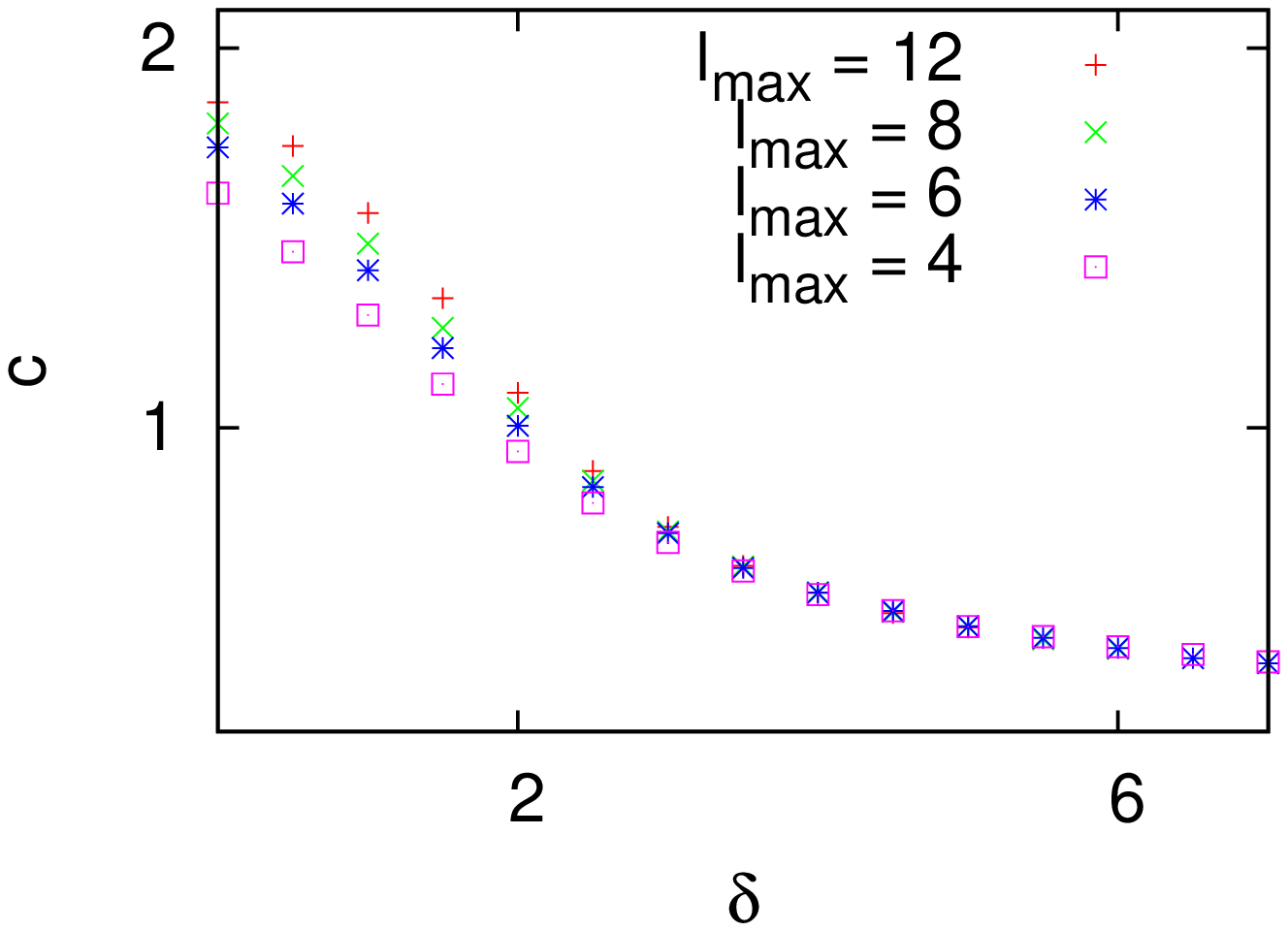}
\includegraphics[angle=0,width=5cm, angle = 270]{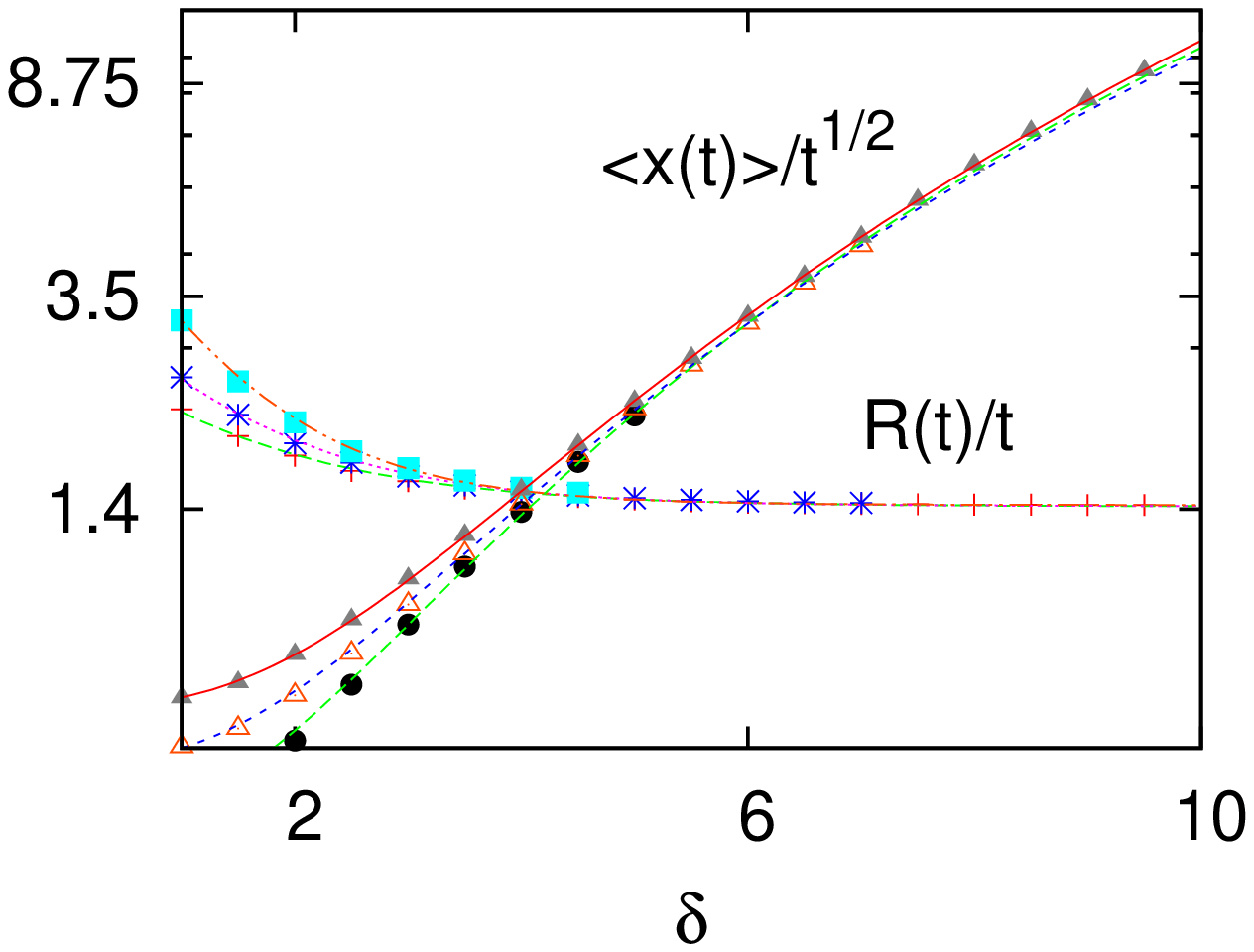}
\caption{Panels (a-c) show that the values of $a,b,c$ (Eq. \ref{eq:scaling}) as function of $\delta$  become independent of $\ell_{max}$ beyond
$\delta \approx 0.4$. Panel (d) shows 
the average distance $\la x\ra$ scaled by $\sqrt{t}$ 
%the time of observation increases with $\delta$ while
and the range of the distribution scaled by $t$ in a log-linear plot.  
%approaches a saturation value
%$\sqrt{2}$. 
The data are shown for $\ell_{max} = 4,6$ and 12 (larger values
of the range correspond to larger $\ell_{max}$ while it is the opposite for  $\la x\ra$)  along with the best fit curves (see text).}
\label{fig4:abc}
\end{figure}

The values of $a,b,c$ are estimated for several values of $\ell_{max}$ and 
are shown as functions of $\delta$  in Fig. \ref{fig4:abc}; 
all these  quantities 
become independent of $\ell_{max}$ beyond $\delta \approx 4$. 

The above results indicate  that there exists a  crossover behaviour 
for the scaling function $g(z)$ at $\delta =\delta^* \approx 4.0$, marked by two features.  First, we find that the  power 
law decay region of the scaling function $g(z)$ vanishes above this value of $\delta$. Secondly, beyond $\delta^*$, 
 $g(z)$    becomes independent  of $\ell_{max}$  ($\geq  4$)  indicating universal behaviour beyond $\delta^*$ with respect to the cutoff.

%vanishing  of the power law 
%region of the scaling function $g(z)$ and also by the fact that it becomes $l_{max}$ independent beyond $\delta^*$. 

In addition,  we  study  two other  quantities which depend on $\delta$ 
and also bear  the signature of the crossover.
 The 
scaled distance $\la x \ra/\sqrt{t} $  travelled at time $t$ 
%travelled  at   time $t_{max}(\delta,\ell_{max}$ up to which the walk is generated is calculated  
%shows that its values   merge beyond $\delta^* \approx4$ for different values of $\ell_{max}$. 
%The variation of  $\la x \ra/\sqrt{t}$ 
is found to be of the form 
\begin{equation}
\la x \ra/\sqrt{t} =  \beta_0 + \beta \delta^\kappa,
\label{average-x}
\end{equation}
%
%(positive values only - negative side, $\beta_0$ slightly less - check!)
where the value of $\kappa$ is  $\approx 2.55 \pm 0.03$ for $\ell_{max} > 4$ and $2.60 \pm 0.01$ for $\ell_{max} = 4$. The exponent value is obtained by fitting the data for $\delta \geq 0.5$.   We conclude that $\kappa $ is a universal 
exponent  $\approx 2.6$ 
independent of $\ell_{max}$ as the values differ by less than ten percent.
%The results are shown in  \ref{fig5:distrange}. 

%We note that 
%$f(x,t)$   becomes more and more asymmetric as $\delta$ is increased (note that for $\delta \to \infty$, when the model becomes effectively  short ranged and  deterministic, one has an asymmetric form for $f(x,t)$). 
%A straightforward measure of the asymmetry may be the value of $\la x\ra$ compared at the same time for different values of $\delta$ for a given $\ell_{max}$. 
%Although it increases  monotonically with  $\delta$, 
% even for a large value of $\delta$, the corresponding result for the QWWOD is much larger, eg., at $t=10000$, $\la x \ra = \simeq 816$ while it is $\simeq 1786$ for the QWWOD. 
%For $\delta > \delta^*$, the values do not depend on $\ell_{max}$, so this 
%is true for any value of the cutoff. 
%On the other hand, for $\delta < \delta^*$, the walk is more restricted and more so  
%for a larger value of $\ell_{max}$ and therefore   $\la x\ra$ is much less 
%compared to the QWWOD value again for any value of $\ell_{max}$. 

The range  $R$ over which the probability distribution is nonzero ($ > 10^{-12}$ numerically) is also calculated. Since the ballistic peaks occur at 
values of $x \sim t$,  the time up to which the  walk is generated,  $R$ is expected to scale as $t$. Hence $R$  is  scaled  by $t$  to 
find out whether any universal behaviour exists. It is found that $R/t$ has the form 
\begin{equation}
R/t = s + q\exp(-r \delta).
\end{equation}
The value of $r$ shows an approximate linear increase with  $\ell_{max}$
while $s$ is very close to   $\sqrt{2}$  independent of $\ell_{max}$. Note that 
for the QWWOD, the peaks occur at $x \approx t/\sqrt{2}$ such that the range at large times is approximately $\sqrt{2} t$ and the scaled range is $\sqrt{2}$ which coincides with the value  obtained here for $\delta \to \infty$. 
The results for $\la x \ra/t^{1/2}$ and $R/t$ are also shown in  Fig. \ref{fig4:abc}.

Both  $\la x \ra /t^{1/2} $ and  $R/t$ derived from the probability distribution $f(x,t)$ 
show the signals of a crossover behaviour as expected. 
However, we find that the result for the QWWOD limit is attained only by  $R/t$ as it saturates to the value $\sqrt{2}$ quite fast beyond $\delta^*$.  
This is due to the fact that the central hump, present even at large values of $\delta$,
affects all  other quantities but not  the range.

The steep rise of $\la x \ra$ with $\delta$ can be understood qualitatively. 
It is  noted that $f(x,t)$ becomes more asymmetric as $\delta$ is increased 
(note that for $\delta \to \infty$, when the model becomes effectively  short ranged and  deterministic, one has an asymmetric form for $f(x,t)$ with the chosen initial condition).  This is clearly indicated by the numerical values of $\la x \ra$ which increase with $\delta$ 
according to Eq. (\ref{average-x}).  Even with  $\delta$ as large as 9, 
 $\la x \ra  \simeq 816$ with $\ell_{max}=4$ while it is $\simeq 1786$ for the QWWOD. 
Apart from the fact that the asymmetry increases with $\delta$, another reason for $\la x \ra$  to increase with $\delta$ is that the height of the central peak
decreases with $\delta$. Thus the  contribution from larger values of $x$ 
become more significant in the expectation value of $x$,   
justifying  the  rather fast non linear increase of $\la x \ra$ with $\delta$.

\begin{figure}
\includegraphics[angle=0,width=4cm, height=5.9cm,angle = 270]{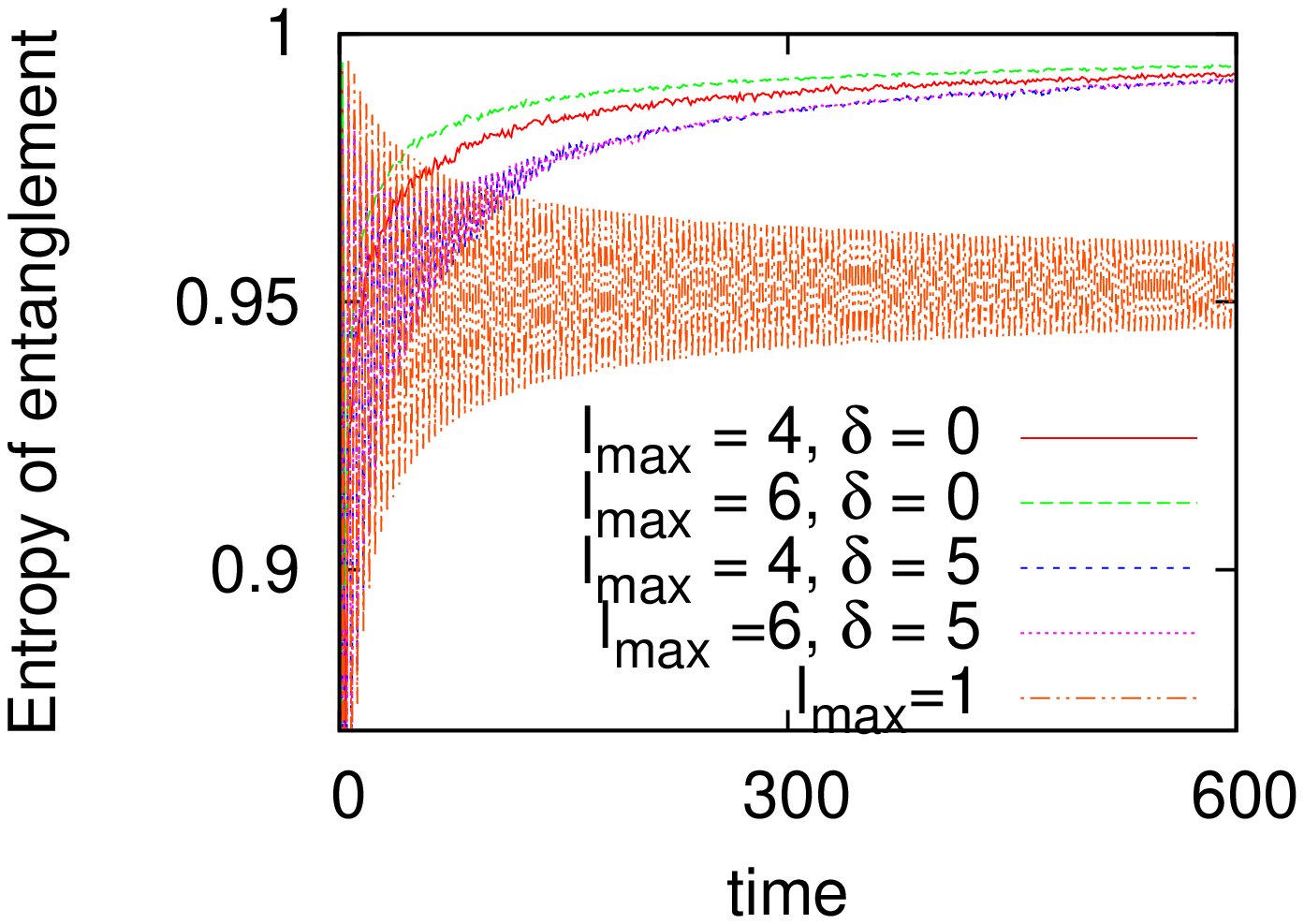}
\includegraphics[angle=0,width=4cm, height=5.8cm,angle = 270]{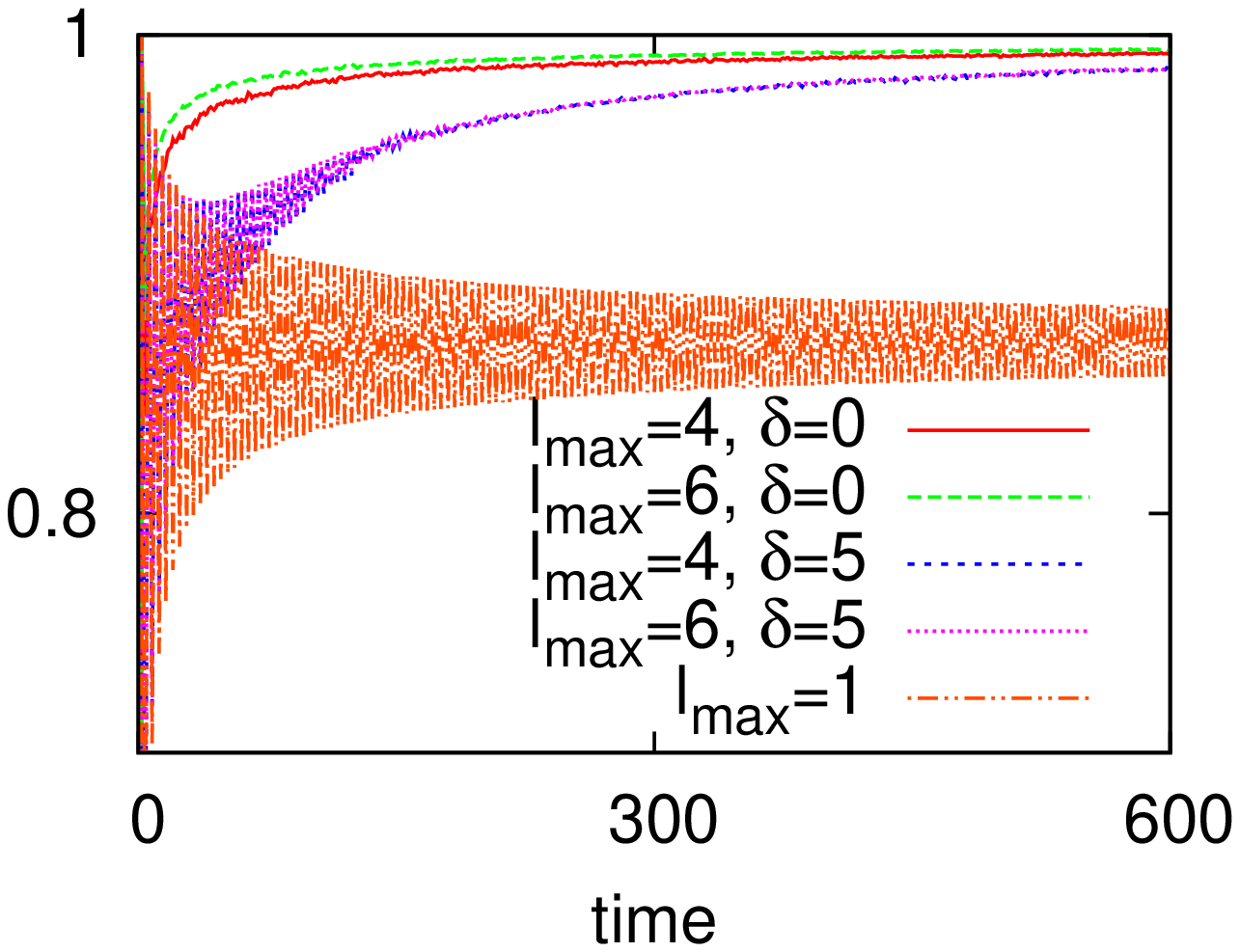}
\caption{The von Neumann entropy of the quantum walk with disorder (QWWD) 
is compared with that of the quantum walk without disorder (QWWOD) for two values of $\delta$ and $\ell_{max}$. The left panel shows the result for the 
initial state $a_0 = 1/\sqrt{3}, b_0 = \sqrt{2/3}$ while for the right panel, the initial state is $a_0 = 0, b_0=1$.  
}
\label{fig6:entangle}
\end{figure}

Lastly, we investigate the so called quantumness of the process by looking at the entanglement measure for different values of the parameters.
We calculate the von Neumann entropy of the reduced density matrix by taking trace over the position variables \cite{ide,abal,carn}. 
Note that the initial state is $|0\ra \otimes (a_0|L\ra + b_0|R\ra)$ 
for which  the von Neumann entropy is zero  but
the entanglement between the position and the coin space develops as the walk 
progresses.
The entropy of the QWWOD is known to be dependent on the initial 
configuration. 
Since the asymptotic entanglement 
is already quite large for the QWWOD with the initial condition used, we show the results for another  initial condition ($a_0=1,b_0 =0$)
for which the asymptotic value is less. 
It is seen that 
the entropy of entanglement   for the QWWD, even at large values of $\delta$, are 
very close to 1,
independent of the initial configuration, and as expected,  
 larger than the QWWOD. 
On the other hand,  we note that
the oscillations are less in amplitude which indicates 
that the interference effects are less, consistent with the existence 
of the central hump occurring even for large (finite) values of $\delta$. 
We show in Fig. \ref{fig6:entangle}  the entropy of the long ranged walk in
comparison to the QWWOD.  As expected, we find that for $\delta = 5$, which is larger than $\delta^*$, the results are independent of $\ell_{max}$ for both the initial conditions. 

%Lastly, we investigate the so called quantumness of the process by looking at the entanglement measure for different values of the parameters.
%We calculate the von Neumann entropy of the reduced density matrix by taking trace over the position variables \cite{}. The entropy of the 
%QWWOD is dependent on the initial configuration \cite{}. Hence we show

\section{Summary and conclusions}

The present work had two aspects, first the effect of long range steps taken from a fat tailed distribution and
second the truncation of the maximum step length. As obtained in the previous studies of random long ranged quantum walks \cite{ps2018,sreetama},
one finds $\la x^2 \ra \propto t^{3/2}$. This further strengthens the claim that the introduction of the long ranged
steps modifies the scaling behaviour in a manner independent of the nature of the distribution $P(\ell)$.

It may be naively thought that for a choice of initial values of $a_0$ and $b_0$ which leads to a asymmetric form of $f(x,t)$,
$\la x \ra$ will scale as $\sqrt{\la x^2\ra} \propto t^{3/4}$. However we find
$\la x \ra$ behaves as $t^{1/2}$.
One can  justify this   using
  the empirical analysis; $f(x,t)$ shows scaling behaviour with $x/t^{1/2}$ as the scaling variable (almost up to the point of  the advent
of the ballistic peaks) and since the contribution from the ballistic peaks
nearly cancel out, the scaling behaviour is determined by the central region of $f(x,t)$ only.

The scaling function has been analyzed in some detail and it shows an interesting crossover behaviour. It may be recalled that the classical
Levy walk shows a crossover to the short ranged behaviour (Gaussian form for probability distribution) beyond a certain value of the parameter $\delta$.
Here, however, we do not get a crossover to the short ranged behaviour (which seems to be present only for $\delta \to \infty$ for $\ell_{max} > 1$) but rather a change in the behaviour of the scaling function and also a universal behaviour with respect to $\ell_{max}$ for $\delta > \delta^*$.
On the other hand the exponents are robust with respect to $\delta$, even for the largest value of $\delta$ considered.
 This may appear counter-intuitive as the system is expected to have short range behaviour for large values of $\delta$ but
apparently, the peak at the center persists even for the maximum  value of
$\delta$ that could be simulated indicating that the interference effects  are less effective compared  to
 the QWWOD.

One may try to infer what happens if $\ell_{max}$ is made infinite, i.e., the unrestricted  Levy walk.
The present results indicate that above $\delta^*$, the scaling function
has a stretched exponential behaviour that is independent of  $\ell_{max}$.
Hence it can be  conjectured that  the scaling function will have a stretched exponential behaviour for $\ell_{max} \to \infty$ as well for $\delta > \delta^*$. However, this needs more investigation since
the results found here are all for finite values of $\ell_{max}$.
One other interesting aspect of the long ranged walk is in the context of searching. We note that with $\delta < \delta^*$, the range is larger compared to
that of the usual walker. So even though the walk is partially localised, it is possible to reach remote targets, the more so with larger $\ell_{max}$,    albeit at a sub-ballistic speed.

When long term memory along with variable step lengths is  considered, the variance  shows a super-ballistic behaviour \cite{elephant}.
Hence from the present results, one can say it is  the effect of memory that makes the walk faster  and not the choice of
time dependent step lengths.

One last point to be mentioned is the issue of quantumness. There have been some very recent works which suggest alternative ways of estimating this \cite{qness1,qness2} in quantum walks, these  can be used in future studies.

\medskip

Acknowledgement: 
The author acknowledges financial support from SERB project EMR/2016/005429.
Discussions with Amit Kumar Pal and Suchetana Mukhopadhyaya are gratefully acknowledged.

\end{document}